\documentstyle[12pt,psfig]{article}
\newcommand{\gray}{$\gamma$-ray}
\newcommand{\grays}{$\gamma$-rays}
\newcommand{\etal}{{\it et al.}}

\newcommand{\apj}{{\it Astrophys. J.}}
\newcommand{\apjl}{{\it Astrophys. J. (Lett.)}}

\newcommand{\xxiicrc}{{\it 21st Internat. Cosmic Ray Conf.}}
\newcommand{\xxvicrc}{{\it 25th Internat. Cosmic Ray Conf.}}
\newcommand{\mic}{$\mu$m}

\renewcommand\figurename{\it Fig.}

\renewcommand\thesection{\arabic{section}.}
\renewcommand\thesubsection{\thesection\arabic{subsection}.}
\renewcommand\thesubsubsection{\thesubsection\arabic{subsubsection}.}
\renewcommand\section[1]{\vspace{\topsep}\vspace{\partopsep}
\refstepcounter{section}

{\par  \noindent\normalsize\bf \thesection
\hspace{1em}#1\vspace{\topsep}\par\noindent}}

\newenvironment{refs}
{\vspace{\topsep}\vspace{\partopsep}
{\par \noindent\normalsize\bf  References
\vspace{-\topsep}\par\noindent}
\setlength{\parindent}{-5mm}
\begin{list}{}{\topsep 0pt \partopsep 0pt \itemsep 0pt \leftmargin 5mm
\parsep 0pt \itemindent -5mm}}
{\end{list}}

\renewcommand\subsection[1]{
\refstepcounter{subsection}
{\par \protect\vspace{\topsep}\vspace{\partopsep}
 \noindent\normalsize\bf \it \thesubsection
\hspace{1em}#1\par \noindent}}

\renewcommand\subsubsection[1]{
\refstepcounter{subsubsection}
{\par \protect \vspace{\topsep}\vspace{\partopsep}
\noindent\normalsize \it \thesubsubsection
\hspace{1em}#1\par \noindent}}

\newfont{\sansb}{cmssbx10}
\newfont{\sans}{cmss10}

\newcommand\tensor[1]{\mbox{\sansb #1}}
\newcommand\vect[1]{\mbox{\bf #1}}
\newcommand\dd{\partial}
\newcommand\sun[0]{\odot}

\begin{document}
\begin{center}
{\large \bf Absorption of Intergalactic TeV Gamma-Rays\vspace{18pt}\\}
{F. W. Stecker$^1$ \& O.C. De Jager$^2$\vspace{12pt}\\}
{\sl
$^1$Laboratory for High Energy Astrophysics, NASA Goddard Space Flight Center, Greenbelt, MD 20771, USA\\
$^2$Space Research Unit, Potchefstroom University, Potchefstroom 2520, 
South Africa\\}
\end{center}

\begin{abstract}
We discuss the problem of the absorption of very high-energy \grays\ by
pair production interactions with extragalactic photons which originate
from stellar emission in the infrared, optical and ultraviolet
and dust absorption of starlight with reradiation in the mid- and far-infrared.
The absorption of \grays\ above 1 TeV is dominated by interactions with 
infrared photons. We make a new determination of the optical depth of the
universe to multi-TeV photons as a function of energy and redshift and use the
results to compare with recent spectral data of Mrk 421 and Mrk 501, sources
that have been observed in the flaring state up to $\sim$10 TeV. For the
optical depth calculations, we have made use of a new, {\it empirically based}
calculation of the intergalactic infrared radiation field (IIRF) which we
consider to be more accurate than that based on previous theoretical modeling. 
We also discuss the absorption of sub-TeV \grays\ by starlight photons at
high redshifts.
\end{abstract}
\setlength{\parindent}{1cm}
\section{Introduction}
The magnitude and shape of the intergalactic infrared radiation field
(IIRF) are fundamental to understanding the 
early evolution of galaxies. 
The IIRF is produced by the emission of radiation
from the photospheres of cool stars, dust reradiation, and redshifting to 
even longer wavelengths. The IIRF is especially sensitive to the 
``instantaneous'' rates of star formation at a given redshift, 
since much of the energy in bursts of new star formation escapes 
in the infrared after being reradiated by dust.
The study of the IIRF is therefore of great cosmological significance.

We have previously pointed out (Stecker, De Jager and Salamon 1992 (SDS92)) 
that very high energy \gray\ beams from blazars can be used to measure  the
intergalactic infrared radiation field, since 
pair-production interactions of \grays\ with intergalactic IR photons 
will attenuate the high-energy ends of blazar 
spectra. Determining the intergalactic IR field, in turn, allows us to 
model the evolution of the galaxies which produce it. 
As energy thresholds are lowered 
in both existing and planned ground-based
air Cherenkov light detectors (Cresti 1996), cutoffs in the \gray\ spectra of 
more distant blazars are expected, owing to extinction by the low energy
stellar photons. These can be used to explore the redshift dependence 
of the stellar emissivity. 
Furthermore, by using blazars for a determination 
of attenuation as a function of
redshift,  combined with a direct observation of the IR background from
the {\it DIRBE}  detector on {\it COBE}, one can, in principle, 
measure of the Hubble
constant $H_{0}$ at truly cosmological distances (Salamon, Stecker and
De Jager 1994). 

The potential importance of the photon-photon pair-production process in 
high energy astrophysics was first pointed out by Nikishov (1961). 
Unfortunately, his early paper overestimated the energy density of the  
IIRF by about three orders of magnitude. However, with the discovery 
of the cosmic microwave background radiation, Jelley (1966) and 
Gould and Schreder (1967) were quick to point out the opacity 
of the universe to photons 
of energy greater than 100 TeV. Stecker (1969) and Fazio and Stecker (1970) 
generalized these calculations to high redshifts, showing that photons 
originating at a redshift $z$ will be absorbed above an energy of $ \sim 
100 (1+z)^{-2}$ TeV. 

There are now over 50 grazars which have been detected by the {\it EGRET} team
(Thompson, \etal\ 1996). These sources, optically violent variable quasars
and BL Lac objects, have been detected out to a redshift greater that 2.
An {\it EGRET}--strength ``grazar'' (observed \gray\
blazar) with a hard spectrum extending to multi-TeV energies is potentially
detectable with ground-based telescopes (see the review on the
Atmospheric Cherenkov Technique by 
Weekes (1988)). Pair-production interactions with
the cosmic microwave background radiation (CBR) will not cut off the \gray\
spectrum of a low-redshift source below an energy of $\sim 100$
TeV (see above). However, as pointed out in SDS92, even 
bright blazars at moderate redshifts ($z >0.1$) will suffer absorption at TeV
energies owing to interactions with the IIRF.

Shortly after SDS92 was published, the Whipple Observatory group reported 
the discovery of $\sim$ TeV \grays\ coming from the blazar Mrk 421, a
BL Lac object with a redshift of 0.031 (Punch \etal\ 1992). This 
source has a hard, roughly $E^{-2}$,
\gray\ spectrum extending to the TeV energy range (Mohanty, \etal\ 1993; 
Lin, \etal\ 1994) 
where the pair-production process involving infrared
photons becomes relevant. Following this discovery, we made use
of the TeV observations of Mrk 421 to carry out our suggestion of using
such sources to probe the IIRF (Stecker and De Jager 1993;
De Jager, Stecker and Salamon 1994). This problem was also 
discussed by Dwek and Slavin (1994) and Biller, \etal\ (1995). 
The TeV spectrum of Mrk 421 obtained by the Whipple Observatory group 
(Mohanty \etal\ 1993) 
showed no significant
absorption out to an energy of at least 3 TeV.
It is this lack of absorption below 3 TeV which we (Stecker and De Jager 1993)
used to put upper limits on the extragalactic infrared energy density in
the $\sim$ 1 $\mu$m to 8 $\mu$m wavelength range (see
also Dwek and Slavin 1994). 

Our upper limits on the IIRF, obtained from the Whipple data
on Mrk 421 below 3 TeV, ruled out various exotic mechanisms for producing
larger fluxes, such as decaying particles, exploding stars, massive
object and black hole models (Carr 1988). They are consistent with the
extragalactic near infrared background originating from ordinary stellar
processes in galaxies (Stecker, Puget and Fazio 1977; 
Franceschini \etal\ 1994; see also the papers included in Dwek 1996). 

The paper by De Jager, \etal\ (1993) attempted to estimate the IIRF based on
prelimary data on the spectrum of Mrk421 given by Mohanty, \etal\ (1993), data
which indicated that there may be absorption above 3 TeV in that source. 
Recent observations of Mrk421 in the flaring state by the same group 
with much better statistics indicate no absorption or weak absorption out
to an energy of at least $\sim$8 TeV (McEnery, \etal\ 1997); 
Thus, the analysis of De Jager, \etal\ (1993) no longer holds. 
We examine the implications of the new Mrk421 data later in this paper.

Of all of the blazars detected by {\it EGRET}, only the low-redshift 
BL Lac, Mrk 421, has been seen by
the Whipple telescope. The fact that the Whipple team did not detect the
much brighter {\it EGRET} source, 3C279, at TeV energies (Vacanti, \etal\ 1990,
Kerrick, \etal\ 1993) is consistent with the predictions of a
cutoff for a source at its much higher redshift of 0.54 (see SDS92).
So too is the recent observation of two other very close BL Lacs ($z < 0.05$),
{\it viz.}, Mrk 501 (Quinn, \etal\ 1996) and 1ES2344+514 (Catanese, 
\etal\ 1997) which were too faint at GeV energies to be seen by {\it EGRET}.

In this paper, we calculate the absorption coefficient of intergalactic
space using a new, empirically based calculation
of the spectral energy distribution (SED) of intergalactic low energy 
photons (Malkan \& Stecker 1998) 
obtained by integrating luminosity dependent infrared spectra of galaxies
over their luminosity distribution (as given by their 60 $\mu$m luminosity 
function) and over their redshift distribution.
After giving our results on the \gray\ optical depth as a function of energy 
and redshift out to a redshift of 0.3, we apply our calculations 
to the high-energy \gray\ spectra of Mrk 421 and Mrk 501 and 
compare our results with recent spectral data on Mrk 421 as reported by 
McEnery, \etal\ (1997) and spectral data on Mrk 501 given by Aharonian,
\etal\ (1997). The results presented here supercede those of our 
previous calculations (Stecker \& De Jager 1997), which were based 
more on theoretical models (see discussion in Malkan \& Stecker 1998).

\section{The Intergalactic Radiation Field}

A theoretical upper limit to the intergalactic infrared energy density 
whose ultimate origin is stellar nucleosynthesis was given by Stecker \etal\
(1977). Direct observational limits on the IIRF, determined by
analysis of the {\it COBE/DIRBE} data with other components 
subtracted out, has been given by Hauser (1996). 
Both of these determinations are well below the limits
obtained by Biller, \etal\ (1995). Biller, \etal\ (1995) 
argued that the most conservative estimate
of an upper limit on the IIRF would be obtained by taking the flattest
spectral index for Mrk 421 consistent with {\it EGRET} data at GeV energies
and assuming that the spectrum of Mohanty et al. (1993) is the result
of absorption with an optical depth $\tau > 6$ in the TeV energy range.
The huge IIRF required to obtain this optical depth is inconsistent with the 
{\it COBE/DIRBE} data (Hauser 1996). Also, in order to
produce a power-law absorbed \gray\ spectrum from a different power-law 
source spectrum in the TeV energy range, they had
to employ an unphysical SED which is inconsistent with that produced by
theoretical models of stellar and dust emission from galaxies.

Recently, Puget, \etal\ (1996) have claimed a tentative detection of
the far-infrared background from an analysis of {\it COBE} data at wavelengths
greater than 200 \mic. 
We have obtained an upper limit to the IIRF (Stecker and De Jager 1993)
based on an analysis of the preliminary
spectrum of Mrk 421 at TeV energies obtained by the Whipple group (Mohanty,
\etal\ 1993).
All of these direct and indirect observational 
results can be brought together into a coherent
scenario where the IIRF is assumed to be generated by stellar processes in
galaxies. The resulting IIRF from both starlight and dust reradiation should
then have a two-peaked SED made up of the contributions from all of the
galaxies in the universe. The energy flux of the IIRF is
expected to be roughly within a factor of two of the value of 10 
nWm$^{-2}$sr$^{-1}$.

Most recently, new estimates of the SED for the IIRF were obtained by 
Malkan \& Stecker (1998), based on
the following empirical data: (1) the luminosity dependent infrared SEDs
of galaxies, (2) the 60$\mu$m luminosity function of galaxies and, (3) 
the redshift distribution of galaxies. This empirically based determination, 
is consistent with the limits given by Stecker \& De Jager (1993) and
the determination of Puget, \etal\ (1996). We consider it to be the most
accurate estimate of the IIRF and will use it here to calculate intergalactic
absorption at low redshifts and multi-TeV energies.

Figure 1 shows the spectral energy distribution (SED) of the extragalactic
infrared background from galaxy emission calculated by Malkan \& Stecker
(1998) together with the cosmic microwave background spectrum (CMBR). 
The IR-SED given by the solid curve in Figure 1 assumes evolution 
out to $z=1$, whereas the IR-SED given by the short dashed curve 
assumes evolution out to $z=2$.
Evolution in stellar emissivity is expected to level off or decrease
at redshifts greater than $\sim 1.5$ (Fall, Charlot \& Pei 1996, Madau 1996)
and evidence for this is discussed in Section 5.1. Thus, 
the two curves in Fig. 1 may be considered to be lower and upper 
limits, bounding the expected IR flux (Malkan \& Stecker 1998).

\vskip 2.0truecm
\centerline{\psfig{figure=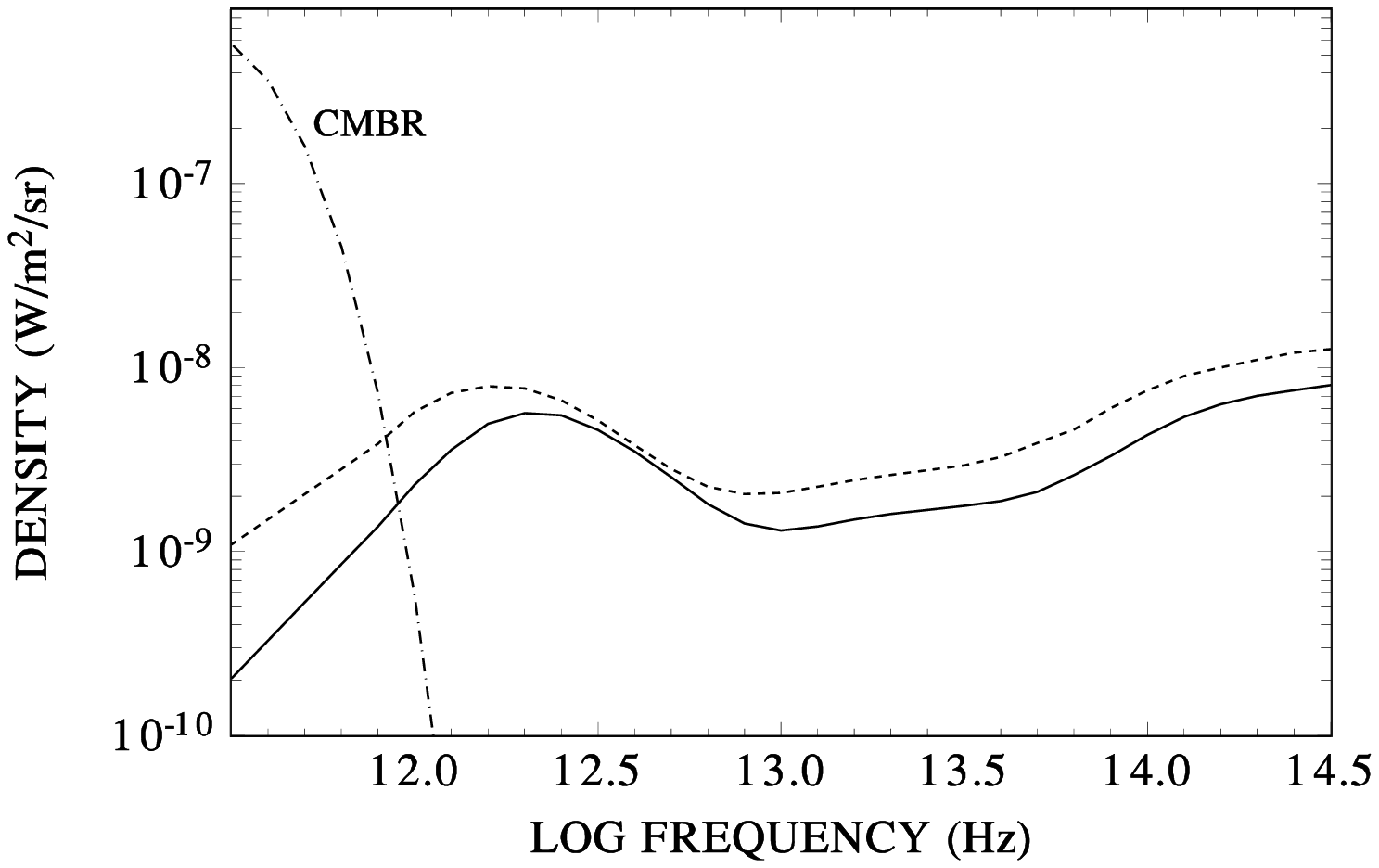,width=13.0truecm}\hskip 0.0truecm}
\vspace{-8.0truecm}
{\smallskip \it Figure 1: The SED of the extragalactic IR background
radiation calculated by Malkan \& Stecker (1998) together with the 2.7K cosmic
microwave background radiation (CMBR). The solid line IR-SED and the short
dashed line IR-SED correspond to the middle and upper curves given by
Malkan \& Stecker with the redshift evolution assumptions as described in
the text.
\smallskip\smallskip}

\section{The Opacity of Intergalactic Space to Low Energy Photons at 
Redshifts $<$ 0.3}

The formulae relevant to calculations involving the pair-production process 
are given and discussed in SDS92.
For $\gamma$-rays in the TeV energy range, the pair-production cross section 
is maximized
when the soft photon energy is in the infrared range: 
$$\lambda (E_{\gamma}) \simeq \lambda_{e}{E_{\gamma}\over{2m_{e}c^{2}}} =
2.4E_{\gamma,TeV} \; \; \mu m \eqno{(1)}$$ where $\lambda_{e} = h/(m_{e}c)$ 
is the Compton wavelength of the electron.
For a 1 TeV $\gamma$-ray, this corresponds to a soft photon having a
wavelength  near the K-band (2.2\mic). (Pair-production interactions actually
take place with photons over a range of wavelengths around the optimal value as
determined by the energy dependence of the cross section.) 
If the emission spectrum of
an extragalactic source extends beyond 20 TeV, then the extragalactic
infrared field should cut off the {\it observed} spectrum between $\sim
20$ GeV and $\sim 20$ TeV, depending on the redshift of the source (Stecker \&
Salamon 1997; Salamon \& Stecker 1998, and this paper).  

The technique for calculating the pair-production optical depth as a 
function of energy and redshift is given in SDS92. In order to calculate the
absorption of multi-TeV \grays\ by IR photons, we
make the reasonable simplifying assumption 
that the IR background is basically in
place at a redshifts $<$ 0.3, having been produced primarily at higher
redshifts (Cowie, \etal\ 1996; Madau 1996; Stecker \& Salamon 1997; 
Salamon \& Stecker 1998 and references therein). 
We therefore limited our calculations to $z<0.3$. For a treatment of 
intergalactic absorption at higher redshifts by optical and UV photons
using recent data on galaxy evolution at moderate and high redshifts,
see Stecker \& Salamon (1997) and Salamon \& Stecker (1998). 
Absorption at higher redshifts is also discussed in Section 5. 

Figure 2 shows the results of our calculations of the optical depth for 
various energies and redshifts up to 0.3.
We take a Hubble constant of $H_o=65$ km s$^{-1}$Mpc$^{-1}$ (Gratton, \etal\
1997). We assume for the IIRF, the two SEDs given by 
Malkan \& Stecker (1997) as shown in Figure 1.

\vskip 2.0truecm
\centerline{\psfig{figure=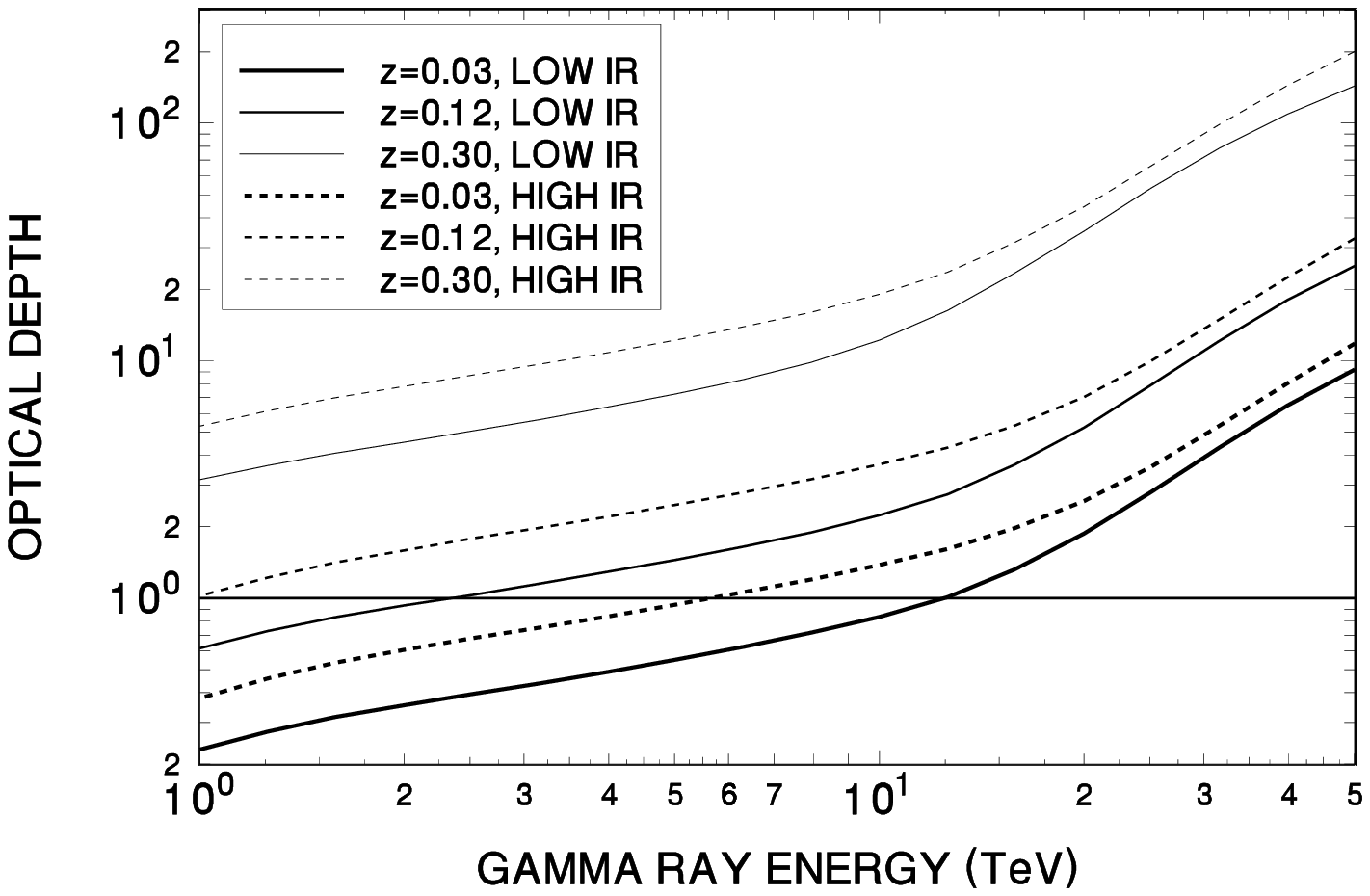,width=13.0truecm}\hskip 0.0truecm}
\vspace{-8.0truecm}
{\smallskip \it Figure 2: Optical depth versus energy for \grays\ originating
at various redshifts obtained using the low IR-SED (solid lines) and 
high IR-SED (short dashed lines) with the IR-SEDs as shown in Figure 1.
\smallskip\smallskip}

Using these two SEDs for the IIRF, we
have obtained parametric expressions for $\tau(E,z)$ for $z<0.3$. The 
double-peaked form of the SED of the IIRF requires
a 3rd order polynomial to reproduce parametrically. It is of the form
$$log_{10}[\tau(E_{\rm TeV},z)]=\sum_{i=0}^3a_i(z)(\log_{10}E_{\rm TeV})^i
\;\;{\rm for}\;\;1.0<E_{\rm TeV}<50, \eqno(2)$$
where the z-dependent coefficients are given by
$$a_i(z)=\sum_{j=0}^2a_{ij}(\log_{10}{z})^{j}. \eqno(3)$$
Table 1 gives the numerical values for 
$a_{ij}$, with $i=0,1,2,3$, and $j=0,1,2$. The numbers before the
brackets are obtained using the lower IIRF SED shown in Figure 1; the
numbers in the brackets are obtained using the higher IIRF SED.
Because we are using real IRAS data to give more accurate estimates of the 
IIRF, we do not give values of $\tau$ for E $<$ 1 TeV
and for larger redshifts, which would involve interactions with 
starlight photons of wavelengths $\lambda \le 1\mu$m (see previous 
discussion and next section). 

\vspace{2em}
\begin{center}
\begin{tabular}{|c|r|r|r|r|}
\multicolumn{5}{c}{\bf Table 1: Polynomial coefficients $a_{ij}$}\\
\hline
$j$ & $a_{0j}$ & $a_{1j}$ & $a_{2j}$ & $a_{3j}$ \\ \hline
0&1.11(1.46) &-0.26(~0.10) &1.17(0.42) &-0.24(~0.07)\\
1&1.15(1.46) &-1.24(-1.03) &2.28(1.66) &-0.88(-0.56)\\
2&0.00(0.15) &-0.41(-0.35) &0.78(0.58) &-0.31(-0.20)\\ \hline
\end{tabular}
\end{center}

\vspace{2em}

Figure 3 shows observed spectra 
for Mrk 421 (McEnery, \etal\ 1997) and Mrk 501 (Aharonian \etal\ 1997) 
in the flaring phase, compared with best-fit spectra of the form
$KE^{-\Gamma}\exp(-\tau)$, with $\tau(E,z)$ given by the two 
SEDs given in Figure 1 and Table 1 and $z$ taken to be 0.03 for both sources. 
Because $\tau < 1$ for $E<10$ 
TeV, there is no obvious curvature in the
{\it differential} spectra below this energy; rather, we obtain a
slight steepening in the power-law spectra of the sources as a result 
of the weak absorption. This result implies that the 
{\it intrinsic} spectra of the sources should be harder by 
$\delta \Gamma \sim 0.2$ in the 
lower IRRF case, and $\sim$ 0.45 in the higher IIRF case. 

\vskip 1.8truecm
\centerline{\psfig{figure=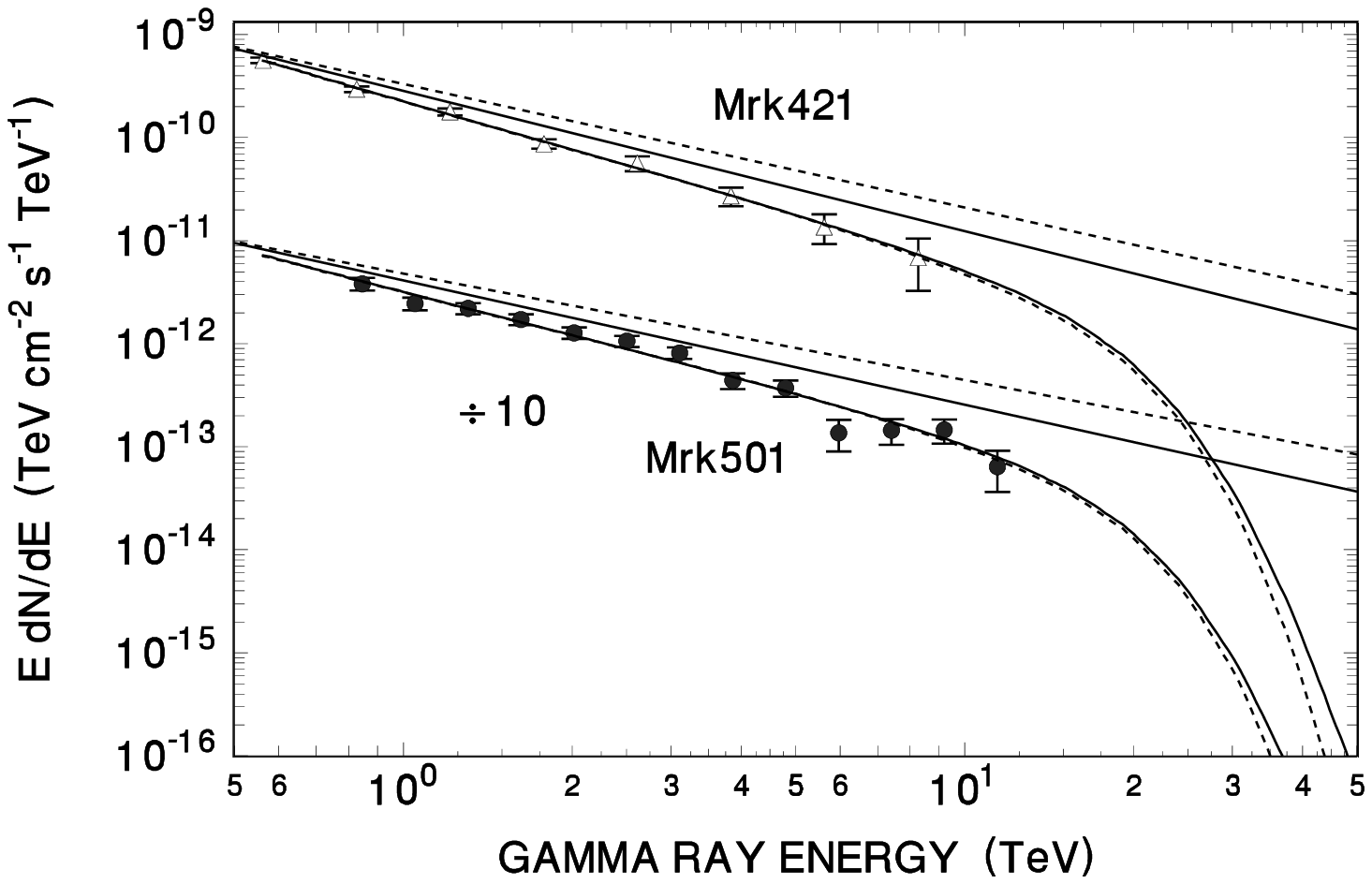,width=13.0truecm}\hskip 0.0truecm}
\vspace{-8.0truecm}
{\smallskip \it Figure 3: The observed spectra of Mrk 421 from McEnery, \etal\
(1997)(open triangles) and Mrk 501 from Aharonian, \etal\ (1997)(solid
circles with the fluxes divided by 10). Best-fit $KE^{-\Gamma}$ power-law 
spectra and spectra with absorption of the form $KE^{-\Gamma}e^{-\tau}$ 
with $\tau(E, z=0.03)$ as shown in Figure 2 are shown for both sources.
For the solid lines corresponding to the low IR-SED, $\Gamma = $2.36 and 2.2 
for Mrk 421 and Mrk 501 respectively. For the short dashed lines corresponding
to the high IR-SED, $\Gamma =$ 2.2 and 2.03 for Mrk 421 and Mrk 501 
respectively.
\smallskip\smallskip}

\section{Reversing the Argument: Constraints on the IIRF from the Gamma-Ray
Observations}

Before closing our discussion of low-redshift sources, it is important to
point out that the recent observations of Mrk 421 and Mrk 501 discussed
above and presented at this conference have accomplished the significant 
feat of providing the best constraints on the IIRF to date, better than
the present direct constraints from the {\it COBE} satellite. By applying 
the same method which we used to give constraints on the IIRF a few years
ago using preliminary data on Mrk 421 (Stecker \& De Jager 1993), we find here
that the present data on Mrk 421 (McEnery, \etal\ 1997) give an upper limit 
to the IIRF at $\sim$ 20 \mic\ of $\sim$ 4 nWm$^{-2}$sr$^{-1}$. A similar 
number has been derived by Stanev \& Franceschini (1997) based on the spectral
data for Mrk 501 of Aharonian, \etal\ (1997). This value is only slightly
higher than the value of 2.6 nWm$^{-2}$sr$^{-1}$ obtained by Malkan \& Stecker
for their upper IR-SED curve as shown in Figure 1. The implication of this
important constraint obtained from very-high-energy \gray\ observations is
very important for understanding the early history of galaxies and star 
formation. It implies that {\it there cannot have been significant evolution
and large amounts of star formation much beyond a redshift of 2.} This 
conclusion is in agreement with very recent {\it HST} and Keck telescope
observations of galaxies at high redshifts as mentioned above and discussed
further in the next section.

\section{Absorption of \grays\ at High Redshifts}
Absorption of $\gamma$-rays from blazars and extragalactic $\gamma$-ray bursts
is strongly dependent on the redshift of the source (SDS92). The study of 
extragalactic absorption below 0.3 TeV at higher redshifts is a complex
and physically interesting subject. In order to calculate such absorption 
properly, one must determine the spectral evolution of galaxy starlight 
photons from the IR through the UV range out to high redshifts.

\subsection{Calculation of the Stellar Emissivity}

Pei \& Fall
(1995) have devised a clever method for calculating stellar emissivity as a 
function of redshift, one which is consistent with all recent data. 
Salamon \& Stecker (1998) (see also Stecker \& Salamon 1997) have calculated
the stellar emissivity at high redshifts by following the method of
Pei \& Fall (1995) and also taking the evolution of stellar metallicity
into account.

The idea of the Pei \& Fall (1995) approach is to 
relate the star formation rate to the evolution of neutral gas density in
damped Ly$\alpha$ systems and then to use the population synthesis models
(Bruzal \& Charlot 1993) 
to calculate the mean volume emissivity of the universe from stars as a 
function of redshift and frequency. Damped Ly$\alpha$ systems are believed
to be either the precursors to galaxies or young galaxies themselves. It is in 
these systems that initial star formation probably took place, so that there is
a relationship between the mass content of stars and gas in these clouds. The
results obtained by Fall, \etal\ (1996) show excellent agreement with 
observational data obtained by the Canada-France redshift survey group for
redshifts out to 1 (Lilly, \etal\ 1996) and are consistent with lower limits
obtained on the emissivity at higher redshifts (Madau 1996). The stellar 
emissivity is found to peak between a redshift of 1 and 2 which is consistent 
with the results of ongoing observations from both the Hubble and Keck 
telescopes. Recent evidence for this peak in the redshift distribution has been
given by Connolly, \etal\ (1997).
 
Salamon \& Stecker (1998) have modified 
the calculations of Fall, \etal\ (1996) by attempting to
account for the significantly lower metallicity of early generation stars at 
higher redshifts which results in increased emission at shorter wavelengths 
and less emission at longer wavelengths. In order to estimate this effect,
they have used the results of Worthey (1994) and moderately extrapolated them
to both lower and higher wavelengths.  They have also
considered the effect of dust opacity and have assumed a reasonable escape
factor to account for the fact that a small fraction of Lyman continuum photons
escape from galaxies unattenuated by stars and dust. The effect of this escape
factor on opacity calculations is negligible, since there are 
not enough ionizing photons in intergalactic space to provide a significant
opacity to multi-GeV \grays.  

\subsection{The Opacity of the Universe at Low-to-High Redshifts}

Once the spectral energy density distribution of stellar photons in 
intergalactic space as a function of redshift is determined, the opacity
of the universe to \grays\ via pair-production as a function of 
\gray\ energy and redshift can
be calculated (SDS92). 
The results obtained by Salamon \& Stecker (1998)
indicate that \grays\ above an energy of $\sim$15 GeV will be attenuated if
they are emitted at redshifts greater than or equal to $\sim$3. The \gray\
burst observed by EGRET on 17 Feb 1994 contained a photon of energy $\sim$18 
GeV. Therefore, it probably originated at a redshift $< 3$. 
Figure 4 shows the calculated opacity as a function of \gray\ 
energy for various source redshifts, with and without the metallicity
correction; the true opacities
likely lie between the values shown in the left and right halves of Fig. 4.

Because the stellar emissivity peaks between a redshift of 1 and 2, there is
little increase in the \gray\ opacity when one goes to redshifts greater
than 2. This weak dependence indicates that the opacity is not determined
by the initial epoch of galaxy formation (at $z \ge 5$), contrary
to the speculation of MacMinn \& Primack (1996).

\vskip 1.0truecm
\centerline{\psfig{figure=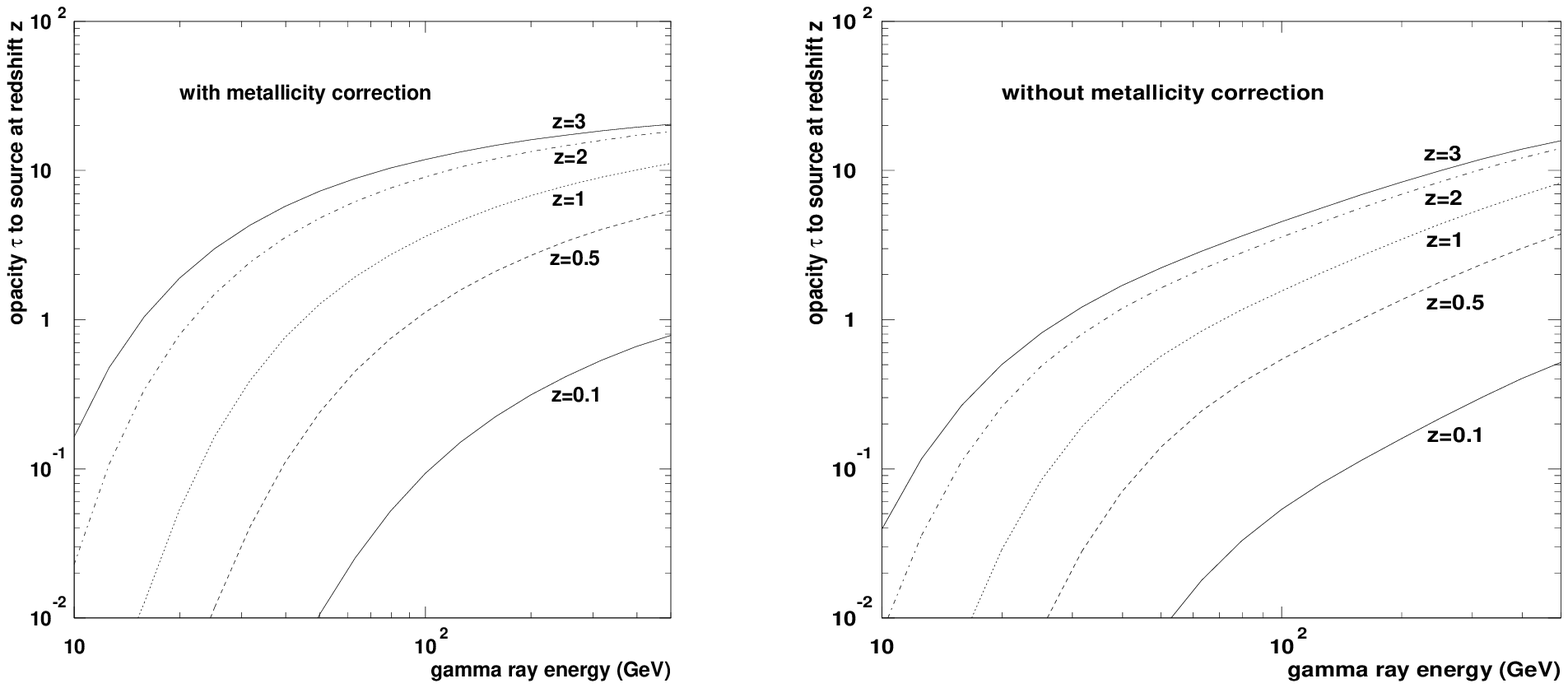,height=8.0truecm}\hskip 0.0truecm}
\vskip-0.0truecm
{\smallskip \it Figure 4: Gamma-ray opacities calculated with and
without metallicity correction factor (Salamon \& Stecker 1998).
\smallskip\smallskip}

\vskip 1.0truecm
\centerline{\psfig{figure=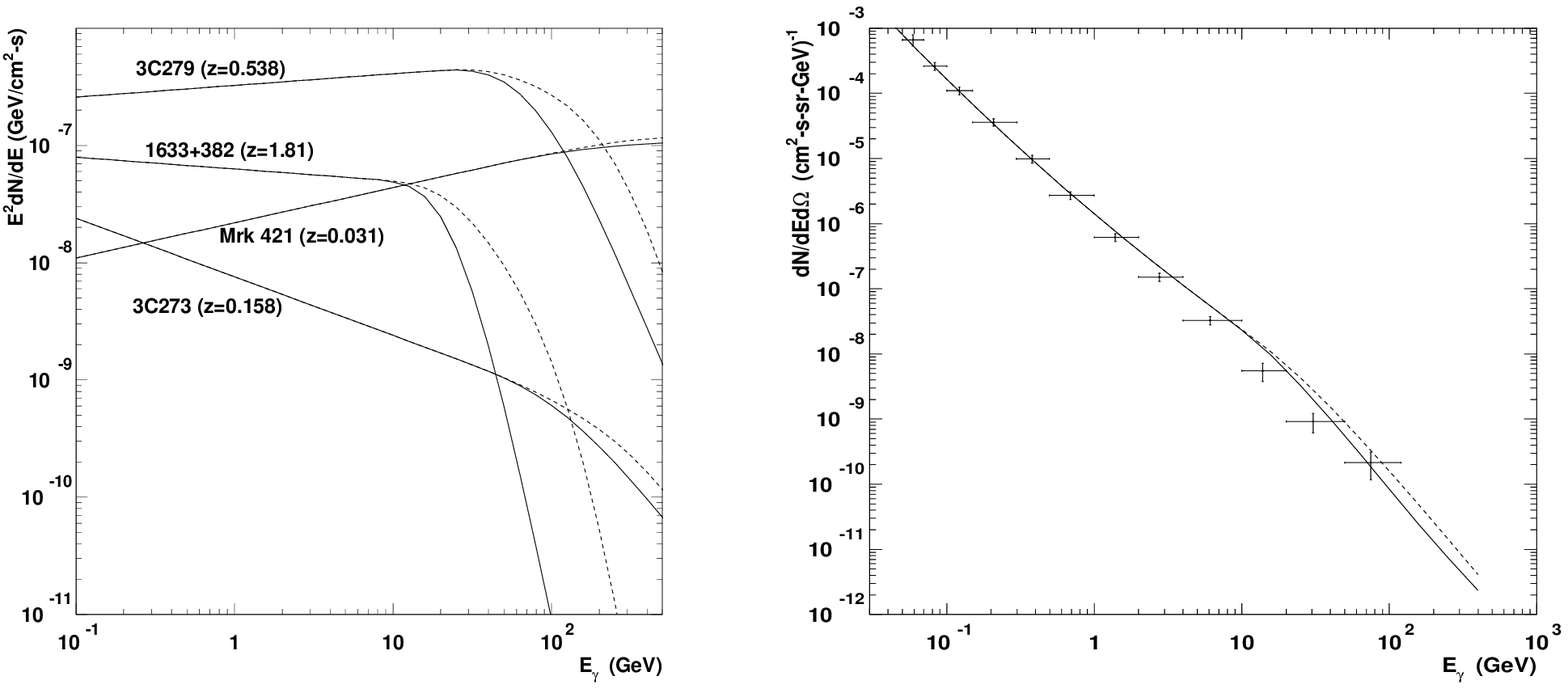,height=9.0truecm}\hskip 0.0truecm}
\vskip-0.0truecm
{\smallskip \it Figure 5: (a) High energy \gray\ power-law spectra of selected 
blazars attenuated
using the opacities given in Figure 4. (b) The extragalactic \gray\ 
background spectrum calculated
with the model of Stecker and Salamon (1996) and attenuated with the
opacities of Fig. 4.   The solid (dashed) lines correspond
to the opacities calculated with (without) the metallicity correction
(Salamon \& Stecker 1998).
\smallskip\smallskip}

\subsection{The Effect of Absorption on Blazar Spectra and the \gray\
Background}

Figure 5a shows the attenuation of \gray\ spectra resulting from
the opacities given in Figure 4 for the blazar sources 1633+382 
($z=1.81$), 3C279 ($z=0.54$), 3C273 ($z=0.16$), and Mrk421 ($z=0.031$).
The solid (dashed) lines result from
the opacities shown in left (right) half of 
Figure 4.  The redshift dependence of the break
energies is evident, as is the absence of any breaks below 10 GeV.
Future measurements of blazar spectral break energies at various redshifts
will thus allow one to discriminate
between cutoffs owing to extragalactic absorption and cutoffs which may be
intrinsic to the source.

Figure 5b shows the effect of absorption on the extragalactic \gray\ 
background computed using the unresolved blazar model of Stecker \& Salamon 
(1996). The solid (dashed) lines correspond to the metallicity correction being
included (neglected) in the opacity calculation.  
The EGRET data on
the extragalactic \gray\ background spectrum (Sreekumar, Stecker and Kappadath
1997; Sreekumar, \etal\ 1998) are also shown in the figure.  

\section{Conclusions}

We have calculated the absorption coefficient of intergalactic space
from interactions with low energy photons of the IIRF,
both as a function of energy and redshift, using 
new estimates of  
SED for the IIRF which were obtained by Malkan \& Stecker (1998).
We have further applied our calculations to the
nearby BL Lac objects Mrk 421 and Mrk 501. The more precise optical 
depth calculations predict that, for these sources, there should not be
a pronounced spectral break
below $\sim 10$ TeV, but rather a small steepening of between 0.2 to
0.45 in the spectral index. The real power 
output from these sources should therefore be
larger than that which would be calculated using the observed spectrum. 
These results are consistent with the recent 
spectral observations of these sources in the flaring state. 

Our new calculations give a smaller intergalactic
absorption effect than we obtained previously (Stecker \& De Jager
1997) because of the lower IIRF calculated by Malkan \& Stecker (1998). The
reasons for this lower predicted IIRF are discussed by Malkan \& Stecker
(1998).

Our calculations predict a significant intergalactic absorption effect
which should cut off the spectra of Mrk 421 and Mrk 501 at energies greater
than $\sim$20 TeV. Observations of these objects at large zenith angles, 
which correspond to large effective threshold energies, may thus demonstrate 
the effect of intergalactic absorption. Cutoffs in the spectra of grazars
at higher redshifts should occur at correspondingly lower energies.

Our calculations confirm the conclusion in SDS92
that the high energy (TeV) spectra of sources 
at redshifts higher than 0.1 should suffer significant absorption. The recent
tentative detection of two additional X-ray selected BL Lac objects (XBLs) at  
redshifts below 0.1, {\it viz.}, 1ES2344+514 (Catanese, \etal\ 1997) and
PKS 2005-489 (K.E. Turver 1997, this conference),
coupled with the non-observations of {\it all} of the known {\it
EGRET} blazars at redshifts above 0.1, even those which are much
brighter than Mrk 421 in the GeV range and have comparable spectral indeces, 
further supports our earlier conclusions regarding the importance
of intergalactic absorption effects in very-high-energy blazar spectra (SDS92)
as well as the argument that nearby XBLs are the only significant TeV 
sources presently detectable (Stecker, De Jager \& Salamon 1996). Indeed,
all of these sources, with estimates of expected flux values, were on our
list of predicted TeV sources (Stecker, De Jager \& Salamon 1996).

In complementary work, Stecker \& Salamon (1997) and Salamon \& Stecker (1998)
have calculated the effect of intergalactic absorption of sub-TeV \grays\ 
by pair production interactions with starlight optical and UV photons
at higher redshifts. This absorption can be important for
energies as low as $\sim$15 GeV at a redshift of $\sim$3 and can have 
a significant effect on the spectra of blazars and \gray\ bursts at 
high redshifts and on the high-energy \gray\ background spectrum. 

\section{Acknowledgments}
We wish to acknowledge R. Lamb, F. Aharonian and G. Hermann for sending us 
their \gray\ data in numerical form.

\begin{refs}

\item Aharonian, F., \etal\ 1997, preprint astro-ph/9706019, {\it
Astron. \& Ap.}, in press

\item Biller, S.D., \etal\ 1995, \apj\ 445, 227

\item Bruzal, A. G. \& Charlot, S. 1993, \apj\ 405, 538

\item Carr, B.J. 1988, in A. Lawrence (ed.),
{\it Comets to Cosmology}, Springer-Verlag: Berlin, p. 265

\item Catanese, M., \etal\ 1997, \xxvicrc, {\it Durban, South Africa} 3, 277

\item Connolly, \etal\ 1997, astro-ph/9706255, \apj, in press


\item[]{} Cowie, L.L. \etal\ 1996, A.J. 112, 839

\item Cresti, M. ed. {\it Towards a Major Atmospheric 
Cherenkov Detector IV}, Padova (1996)

\item De Jager, O.C., Stecker, F.W. \& Salamon, M.H. 1994, 
{\it Nature}\ 369, 294 

\item Dwek, E. 1996, (ed.) {\it Unveiling the Cosmic Infrared Background}, 
AIP CP 348, (New York: Amer. Inst. Phys.) 

\item Dwek, E. \& Slavin, J. 1994, \apj, 436, 696.

\item Fall, S.M., Charlot, S. \& Pei, Y.C. 1996, \apj, 402, 479

\item Fazio, G.G. \& Stecker, F.W. 1970, {\it Nature}\ 226, 135

\item Franceschini, \etal\ 1994, \apj\ 427, 140

\item Gould, R. J. \& Schreder, G.P. 1966, 
{\it Phys. Rev. Letters}\ 16, 252

\item Gratton, R.G., \etal\ 1997, preprint astro-ph/9707107

\item Hauser, M. G. 1996, in 
{\it Unveiling the Cosmic Infrared Background}, AIP CP 348, ed.
E. Dwek, Amer. Inst. Phys.:New York, 11

\item Jelley, J.V. 1966, {\it Phys. Rev. Lett.} 16, 479


\item Kerrick, A.D. \etal\ 1993, 
{\it Proc. 23rd Int'l. Cosmic Ray Conf.},
U. Calgary Press, Calgary, 1, 405

\item Lilly, S.J., Le Fevre, O. Hammer, F. \& Crampton, D. 1996, \apj\ 460, L1

\item Lin, Y.C., \etal\ 1994, in {\it The 2nd Compton Symposium}, 
ed. C. Fichtel, N. Gehrels and J.P. Norris, AIP CP 304, 
Amer. Inst. Phys.:New York, 582

\item MacMinn, D, \& Primack, J. 1966, in {\it TeV Gamma Ray Astrophysics},
ed. H.J. V\"{o}lk and F.A. Aharonian, (Dordrecht: Kluwer) 75, 413

\item Madau, P. 1996, in {\it Star Formation Near and Far}, ed. S.S. Holt \&
L.G. Mundy, AIP Symp. Proc. No. 393 (New York: Amer. Inst. Phys.), 481


\item Malkan, M.A. \& Stecker, F.W. 1998, astro-ph/9710072, \apj, in press

\item McEnery, J.E., \etal\ 1997, \xxvicrc, {\it Durban, South Africa} 3, 257

\item Mohanty, G. \etal\  1993, 
{\it Proc. 23rd Int'l. Cosmic Ray Conf.},
Univ. of Calgary Press, Calgary, 1, 440

\item Nikishov, A.I. 1962, {\it Sov. Phys. JETP}\ 14, 393

\item Pei, Y. C. \& Fall, S. M., 1995, \apj\ 454, 69

\item Puget, J.-L. \etal\ 1996, {\it Astron. and Ap.} 308, L5

\item Punch, M. \etal\ 1992, {\it Nature}\ 358, 477

\item Quinn, \etal\ 1996, \apj, 456, L83

\item Salamon, M.H., Stecker, F.W. \& de Jager, O.C., 1994, \apjl\ 423,
L1 

\item Salamon, M. H. \& Stecker, F. W. 1998, astro-ph/9704166, \apj, in press 
(1 Feb.)

\item Sreekumar, P., Stecker, F.W. \& Kappadath, S.C. 1997, astro-ph/9709258,
{\it Proc. 4th Compton Symp. (Williamsburg, VA)}, in press

\item Sreekumar, P. \etal\ 1996, \apj\ 464, 628

\item Sreekumar, P. \etal\ 1998, astro-ph/9709257, \apj\ 494, in press,
(20 Feb.)

\item Stanev, T. \& Franceschini, A. 1997, preprint astro-ph/9708162

\item Stecker, F.W. 1969, \apj\ 157, 507

\item Stecker, F.W. \& De Jager, O.C. 1993, \apjl\ {\bf 415}, L71 

\item Stecker, F.W. \& De Jager, O.C. 1997, \apj\ 476, 712 

\item Stecker, F.W., Puget, J.L. \& Fazio, G.G. 1977, \apjl\
214, L51 

\item Stecker, F.W., \& Salamon, M.H. 1996, \apj\ 464, 600

\item Stecker, F.W. \& Salamon, M.H. 1997, {\it Proc. 25th Intl. Cosmic Ray
Conf., Durban, South Africa} 3, 317

\item Stecker, F.W., De Jager, O.C. \& Salamon, M.H. 1992,
\apjl\ 390, L49 (SDS92)

\item Stecker, F.W., De Jager, O.C. \& Salamon, M.H. 1996, \apjl\ 473, L75

\item Thompson, D. J., \etal\ 1996, \apj\ {\it Suppl.}, 107, 227

\item Vacanti, G., \etal\ 1990, \xxiicrc\ , 2, 329

\item Weekes, T.C. 1988, {\it Physics Reports}\ 160, 1

\item Worthey, , G. 1994, ApJS, 95, 107

\end{refs}

\end{document}